# Quantum state expansion

Jae-Seung Lee and A. K. Khitrin

Department of Chemistry, Kent State University, Kent, Ohio 44242-0001

**Abstract**

It is experimentally demonstrated that an arbitrary quantum state of a single spin 1/2: $a|\uparrow\rangle + b|\downarrow\rangle$ can be converted into a superposition of the two ferromagnetic states of a spin cluster: $a|\uparrow\uparrow\cdots\uparrow\uparrow\rangle + b|\downarrow\downarrow\cdots\downarrow\downarrow\rangle$. The physical system is a cluster of seven dipolar-coupled nuclear spins of single-labeled $^{13}$C-benzene molecules in a liquid-crystalline matrix. In this complex system, the pseudopure ground state and the required controlled unitary transformations have been implemented. The experimental scheme can be considered as an explicit model of quantum measurement.

Theory of quantum measurements, which describes a boundary between quantum and classical worlds [1,2], is the least established part of quantum theory. Different approaches to this problem lead to different interpretations of quantum mechanics. A serious difficulty in exploring this subject is that practical measuring devices are too complex to allow a detail analysis of their dynamics. It may be helpful to consider some simple and explicit models of quantum measurement, using systems with controllable quantum dynamics. One of possible experimental models is proposed and studied in this work.

Let us consider a system of $N+1$ spins 1/2 (qubits) in the initial state

$$|\psi_{in}\rangle = (a|0\rangle_0 + b|1\rangle_0)|0\rangle_1|0\rangle_2\ldots|0\rangle_{N-1}|0\rangle_N, \qquad |a|^2 + |b|^2 = 1, \qquad (1)$$

where the qubit notations $|0\rangle_k \equiv |\uparrow\rangle_k$ and $|1\rangle_k \equiv |\downarrow\rangle_k$ are used. In this state, the 0-th qubit is in some arbitrary state, defined by two complex coefficients $a$ and $b$, while the qubits 1 to $N$ are in the ground state. Quantum logic circuit [3]

$$U = CNOT_{N-1,N} CNOT_{N-2,N-1} \ldots CNOT_{1,2} CNOT_{0,1} \qquad (2)$$

is a chain of unitary controlled-not gates $CNOT_{m,n}$, which flip the target qubit $n$ when the control qubit $m$ is in the state $|1\rangle_m$ and do not change the qubit $n$ when the qubit $m$ is in the state $|0\rangle_m$. If the 0-th qubit is in the state $|1\rangle_0$, it flips the qubit 1, the qubit 1 flips the qubit 2, and so on. A wave of flipped qubits, triggered by the 0-th qubit, propagates until it covers the entire system. As a result, the circuit (2) converts the initial state (1) into the final state

$$|\psi_{out}\rangle = U|\psi_{in}\rangle = a|0\rangle_0|0\rangle_1\ldots|0\rangle_{N-1}|0\rangle_N + b|1\rangle_0|1\rangle_1\ldots|1\rangle_{N-1}|1\rangle_N. \qquad (3)$$

This state is a superposition of the most macroscopically distinct states: the ferromagnetic states with all spins up and all spins down. An advantage of the circuit (2) is that it requires only interactions between neighbor qubits and, therefore, potentially can be implemented in large systems.

One might think that, since a macroscopic polarization is associated with the state (3), a single measurement can provide some information about the state. However, transforming the state (1) to the state (3) does not decrease relative quantum fluctuations. It can be seen, as an example, by considering an ensemble average of the square of polarization: it has its maximum value, $(N+1)^2$, indicating that possible outcomes of polarization measurement for a single system are one of the two extreme values, $\pm(N+1)$ (Ensemble average values of powers of an observable give unambiguous information about probabilities of possible outcomes in single-system measurement). Therefore, the only goal of creating the state (3) is to increase a signal produced by a system, without making this signal more classical.

The experimental scheme for converting the initial state (1) into the state (3) has been implemented on a cluster of seven dipolar-coupled nuclear spins. The experiment has been performed with a Varian Unity/Inova 500 MHz NMR spectrometer. The sample contained 5% of single-labeled $^{13}$C-benzene (Aldrich) dissolved in liquid-crystalline solvent MLC-6815 (EMD Chemical). In this system, fast molecular motions average out all intermolecular spin-spin interactions. Intramolecular dipole-dipole interactions are not averaged to zero due to orientational order induced by a liquid-crystalline matrix. Therefore, the system is a good example of an ensemble of non-interacting spin clusters, where each benzene molecule contains seven nuclear spins, one $^{13}$C and six protons, coupled by residual dipole-dipole interactions. The spin Hamiltonian is

$$H = -\omega_C I_{0Z} - \omega_H \sum_{k=1}^{6} S_{kZ} + \sum_{k=1}^{6}(b_{0k} + J_{0k})I_{0Z}S_{kZ} + \sum_{k>j>0}^{6} b_{jk}(S_{kZ}S_{jZ} - \frac{1}{2}S_{kX}S_{jX} - \frac{1}{2}S_{kY}S_{jY}), \quad (4)$$

where index 0 is used for the $^{13}$C spin and indexes 1 to 6 numerate $^{1}$H spins, starting from the one closest to $^{13}$C nucleus. $I$ and $S$ are corresponding spin operators, $\omega_C$ and $\omega_H$ are the Larmor frequencies, $b_{jk}$ are the constants of residual dipole-dipole interaction, and $J_{jk}$ are the J-coupling constants. Among J-constants, only $J_{01}$ has a considerable value of $J_{01}/2\pi = 158$ Hz [4], the rest of the J-coupling constants are small and can be neglected on the time scale of our experiment. The $^{1}$H and $^{13}$C thermal equilibrium spectra of $^{13}$C-benzene in MLC-6815 are presented in Figs. 2a and 2a', respectively. For individual peaks in equilibrium spectra, the longitudinal relaxation times (T$_1$) were measured to be 1.7-2.3 sec for proton spins and 1.4-2.9 sec for the $^{13}$C spin. Transverse relaxation times (T$_2$) were measured to be 0.1-0.7 sec for proton spins and 0.3-1.1 sec for the $^{13}$C spin.

In what follows, we will use spin notations $|\uparrow\rangle$ and $|\downarrow\rangle$ for the two states of the $^{13}$C spin, $|u\rangle = |\uparrow\uparrow\uparrow\uparrow\uparrow\uparrow\rangle$ and $|d\rangle = |\downarrow\downarrow\downarrow\downarrow\downarrow\downarrow\rangle$ for the states of protons with all spins up or down.

Two experimental challenges for a seven-qubit system are preparing the state (1) and implementing the unitary operation which converts the state (1) into the state (3). For the system under study, we have recently demonstrated that the superposition state of protons $2^{-1/2}(|u\rangle + |d\rangle)$ could be used to amplify the effect of interaction with $^{13}$C spin [5]. Some basic elements used in that work, together with preparation of the pseudopure state $|u\rangle$ for the proton subsystem [6], are also the building blocks of the experimental scheme in Fig. 1, and will be only briefly described here. More details can be found in Refs. [5] and [6]. To better understand the following steps, it is convenient to introduce the Pauli operators for a subspace of two proton states $|u\rangle$ and $|d\rangle$ : $\Sigma_Z = (1/2)(|u\rangle\langle u| - |d\rangle\langle d|)$, $\Sigma_X = (1/2)(|u\rangle\langle d| + |d\rangle\langle u|)$ and $\Sigma_Y = (i/2)(|u\rangle\langle d| - |d\rangle\langle u|)$.

The first two steps, A and B (Fig. 1) are designed to prepare the pseudopure ground state $|\uparrow\rangle|u\rangle$ of the seven-spin cluster. The experiment starts with a sequence of 90° pulses and gradient pulses to saturate the $^{13}C$ magnetization. Then, the proton magnetization is converted into multiple-quantum coherences by 20 cycles of the eight-pulse sequence [7]. The pulse sequence automatically decouples protons from the $^{13}C$ spin. The six-quantum (6Q) coherence $\Sigma_Y$ is filtered by a combination of phase cycling and 180° pulse. The next step is the evolution caused by the interaction with the $^{13}C$ spin, which rotates $\Sigma_Y$ towards $\pm\Sigma_X$ depending on the state of the $^{13}C$ spin. After 90° rotation, the state density matrix $-I_{0Z}\Sigma_X$ is achieved. Then, evolution with another 20-cycle pulse sequence follows. The multi-pulse period, with about 90% fidelity, corresponds to a "90° pulse" in the $\Sigma$-subspace around some axis in the XY-plane. The phase of this "pulse" in the $\Sigma$-subspace is adjusted by the global phase of the pulse sequence. As an example, 90° phase shift can be achieved by $90°/6 = 15°$ phase shifting of all pulses of the sequence. The "90° Y-pulse" in the $\Sigma$-subspace converts the state $-I_{0Z}\Sigma_X$ into the state $\rho_A = I_{0Z}\Sigma_Z = (|\uparrow\rangle\langle\uparrow| - |\downarrow\rangle\langle\downarrow|)(|u\rangle\langle u| - |d\rangle\langle d|)$. This state is the

mixture of four pure states. The $^1$H and $^{13}$C linear-response spectra for this state are shown in Figs. 2b and 2b', respectively.

The pseudopure ground state $|\uparrow\rangle|u\rangle$ ($\rho_B = |\uparrow\rangle\langle\uparrow||u\rangle\langle u|$) is obtained by redistributing the excess population of the other three states. Redistribution is achieved by partial saturation with Gaussian shaped pulses, which have practically zero spectral intensity at the frequencies of allowed single-quantum transitions from the state $|\uparrow\rangle|u\rangle$ (step B). As a result, population of the state $|\uparrow\rangle|u\rangle$ remains "trapped". The $^1$H and $^{13}$C linear-response spectra for the pseudopure state $|\uparrow\rangle|u\rangle$ are presented in Figs. 2c and 2c', respectively. It may be noted that pseudopure states for systems of up to seven spins have been already demonstrated with liquid-state NMR [8]. However, our seven-spin system with dipole-dipole interactions is dynamically much more complex than any of the previously studied systems. Compared to ZZ-coupled system in Ref. [8], dipole-dipole couplings have all three components, providing more "mixing" dynamics. As a result, the maximum number of peaks in a spectrum of $N$ dipolar coupled spins, $\binom{2N}{N+1} \sim 2^{2N}$, grows much faster with increasing $N$ than the number of peaks in a spectrum of ZZ-coupled system, $N2^{N-1}$. Integrated intensities of the pseudopure state spectra, relative to that of the thermal equilibrium spectra, for proton and carbon spins are 3.5% and 3.3%, respectively. To verify the state, we applied hard 180° pulses on the $^{13}$C and proton spins and compared the spectra with numerically calculated spectra. As one illustration, the spectra for the state $|\downarrow\rangle|u\rangle$ are shown in Figs. 2d and 2d'. The calculated spectra for the states $|\uparrow\rangle|u\rangle$ and $|\downarrow\rangle|u\rangle$ are shown in Figs. 2e and 2e'.

In step C, a $\theta$-pulse on the $^{13}$C spin prepares the state (1) with $a = \cos(\theta/2)$ and $b = \sin(\theta/2)$. To convert this state into the state (3), the same pulse sequence as in step A was used without any phase cycling. The first "90° pulse" in the $\Sigma$-subspace converted the state of proton spins $|u\rangle$ to the entangled state $2^{-1/2}(|u\rangle + |d\rangle)$. Then, interaction delay caused Z-rotation in $\Sigma$-subspace, depending on the state of the $^{13}$C spin, to yield the state $\cos(\theta/2)|\uparrow\rangle(|u\rangle + |d\rangle) + \sin(\theta/2)|\downarrow\rangle(|u\rangle - |d\rangle)$. The relative phase of the last 20-cycle pulse sequence was set to 45°, creating 270° phase shift in the $\Sigma$-subspace. This "90° pulse" in the $\Sigma$-subspace created the final state (3): $\cos(\theta/2)|\uparrow\rangle|u\rangle + \sin(\theta/2)|\downarrow\rangle|d\rangle$. The linear-response spectra for this state at $\theta = 0°$, 90°, and 180° are shown in Fig. 3.

When the $^{13}$C spin was prepared in an eigenstate $|\uparrow\rangle$ ($|\downarrow\rangle$), the seven-spin system resulted in the state with all spins up (all spins down) to give spectra of Fig. 3a and 3a' (Fig. 3c and 3c'). Ratios of peak intensities of the spectra in Figs. 3a and 2c were estimated to be 83.3% and 76.5% for left and right peaks, respectively; the ratio for the $^{13}$C peaks in Figs. 3a' and 2c' is 67.8%. When the $^{13}$C spin was prepared in a superposition state $2^{-1/2}(|\uparrow\rangle + |\downarrow\rangle)$, the spectra in Figs. 3b and 3b' revealed proper correlation between the six proton and $^{13}$C spins.

The implemented experimental scheme can be considered as an explicit model of quantum measurement. Six proton spins represent a measuring device designed to measure a quantum state of the $^{13}$C spin. Conversion of the state (1) to the state (3) takes 7.2 ms. For our small system, this time is considerably shorter than the decoherence time, about 50 ms, of the seven-quantum (7Q) coherence in the state (3). After a time, longer than the 7Q decoherence time but much shorter than $T_1$ (2 sec), two off-diagonal elements of the density matrix of the state (3) decay, and the pure state (3) is converted into a mixed state with the density matrix $|a|^2 |\uparrow\rangle\langle\uparrow| |u\rangle\langle u| + |b|^2 |\downarrow\rangle\langle\downarrow| |d\rangle\langle d|$. This state is indistinguishable from the mixture, with fractions $|a|^2$ and $|b|^2$, of molecules in one of the two pure states: $|\uparrow\rangle|u\rangle$ or $|\downarrow\rangle|d\rangle$. Each of the molecules presents a result of individual measurement, where "macroscopic" magnetization of protons gives the result of this measurement while the state of the $^{13}$C spin is collapsed to the corresponding eigenstate. In the studied experimental model, different dynamic processes associated with quantum measurement have different time scales and can be analyzed separately.

The work was supported by the Kent State University and the US-Israel Binational Science Foundation.

P.S. This manuscript has been submitted to Physical Review Letters on Aug 20, 2004 and was presented at KIAS-KAIST 2004 Workshop on Quantum Information Science on Aug 31, 2004.

**Figure captions**

Fig. 1. NMR pulse sequence.

Fig. 2. (a) and (a') $^1$H and $^{13}$C spectra of the thermal equilibrium state; (b) and (b') $^1$H and $^{13}$C spectra of the state $\rho_A = I_{0Z}\Sigma_Z$; (c) and (c') $^1$H and $^{13}$C spectra of the pseudopure state $|\uparrow\rangle|u\rangle$; (d) and (d') $^1$H and $^{13}$C spectra of the pseudopure state $|\downarrow\rangle|u\rangle$; (e) and (e') numerically calculated $^1$H and $^{13}$C spectra.

Fig. 3. (a) and (a') $^1$H and $^{13}$C spectra at $\theta = 0°$; (b) and (b') $^1$H and $^{13}$C spectra at $\theta = 90°$; (c) and (c') $^1$H and $^{13}$C spectra at $\theta = 180°$.

Fig. 1

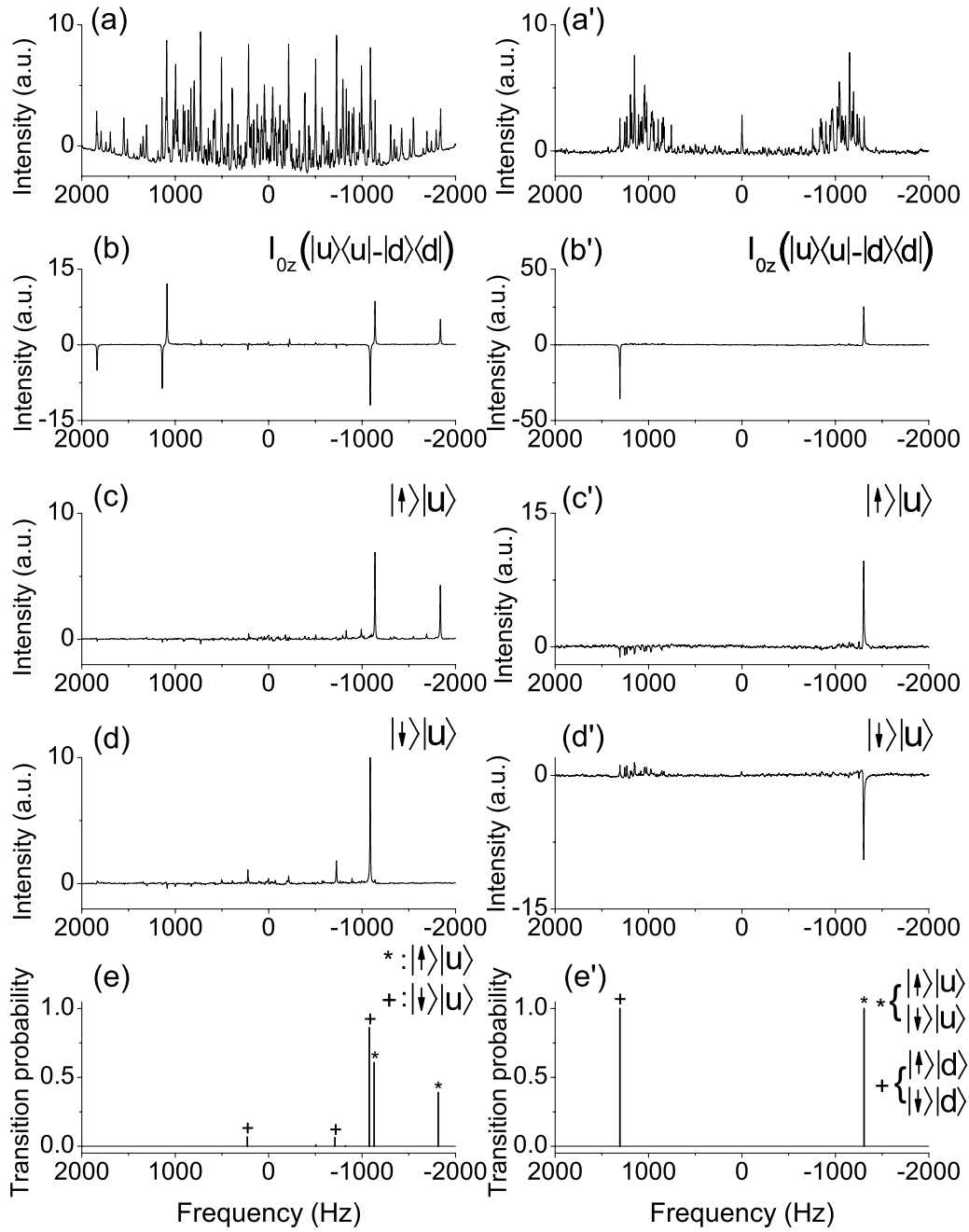

Fig. 2

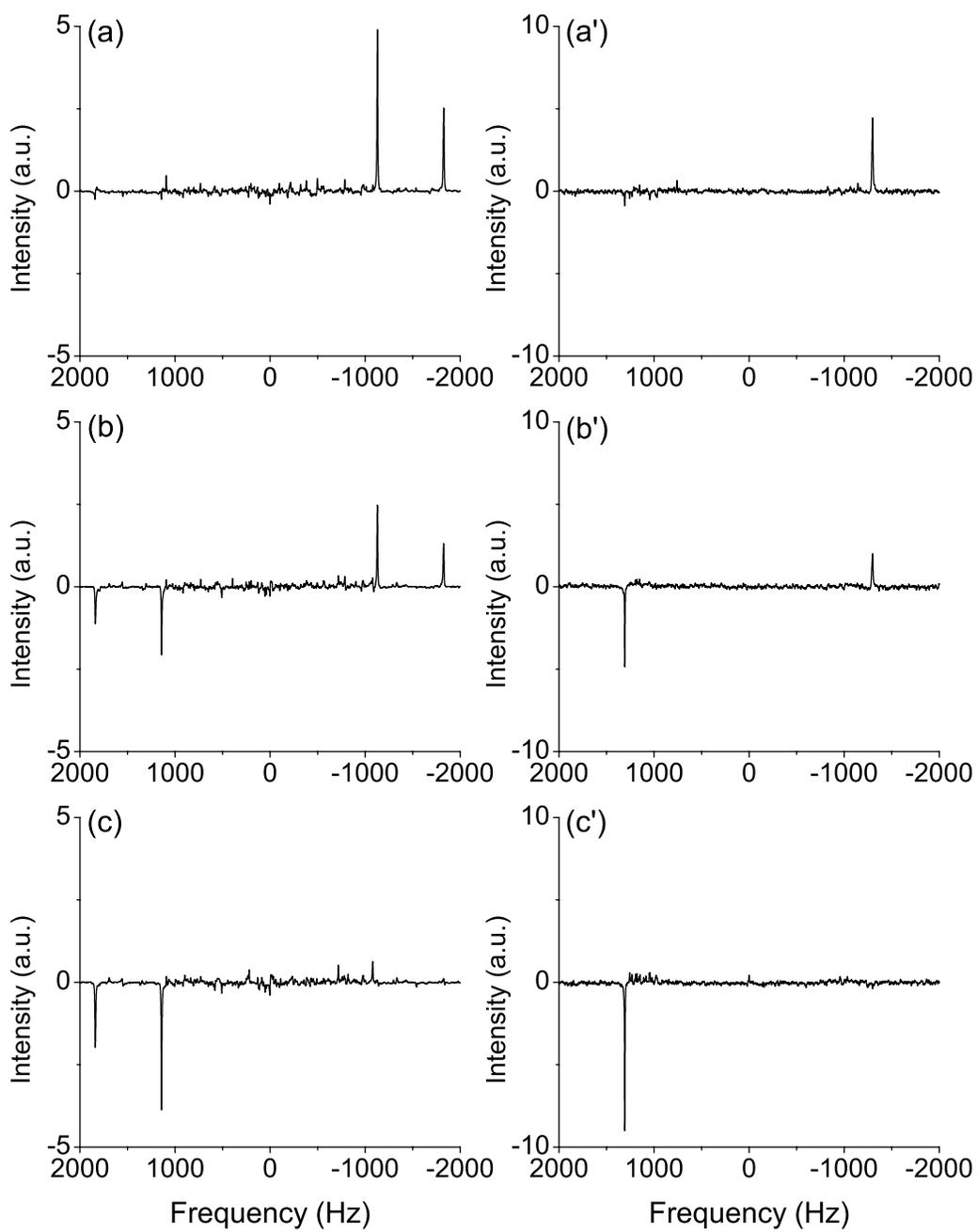

Fig. 3